\documentclass[12pt]{article}
\usepackage{pazha}
\usepackage{graphicx}
\usepackage{lscape}
\usepackage{amssymb}
\usepackage{amsmath} 

\tightenlines
\def\ÐÍ{$\pm$}

\def\deg{$^\circ$}

\begin{document}

{\it ``Astronomy Letters'', 2007, v.33, N7, pp. 455-467}

\bigskip

\title{\bf On the Possibility of Observing the Shapiro Effect for
Pulsars in Globular Clusters}

\author{\bf T.I.Larchenkova\affilmark{1,*} and A.A.Lutovinov\affilmark{2,**}}

\affil{ {\it Astro Space Center, Lebedev Physical Institute, Russian
Academy of Sciences, Profsoyuznaya str., 84/32, Moscow, 117997
Russia}$^1$\\
{\it Space Research Institute, Russian Academy of
Sciences, Profsoyuznaya str., 84/32, Moscow, 117997 Russia}$^2$\\ }
\vspace{2mm}
\received{13 Feb 2007}

\sloppypar
\vspace{2mm}
\noindent

For pulsars in globular clusters, we suggest using observations of the
relativistic time delay of their radiation in the gravitational eld of
a massive body (the Shapiro effect) located close to the line of sight
to detect and identify invisible compact objects and to study the
distribution of both visible and dark matter in globular clusters and
various components of the Galaxy. We have derived the dependences of
the event probability on the Galactic latitude and longitude of
sources for two models of the mass distribution in the Galaxy: the
classical Bahcall-Soneira model and the more recent Dehnen-Binney
model. Using three globular clusters (M15, 47 Tuc, Terzan 5) as an
example, we show that the ratios of the probability of the events due
to the passages of massive Galactic objects close to the line of sight
to the parameter f2 for pulsars in the globular clusters 47 Tuc and
M15 are comparable to those for close passages of massive objects in
the clusters themselves and are considerably higher than those for the
cluster Terzan 5. We have estimated the rates of such events. We have
determined the number of objects near the line of sight toward the
pulsar that can produce a modulation of its pulse arrival times
characteristic of the effect under consideration; the population of
brown dwarfs in the Galactic disk, whose concentration is comparable
to that of the disk stars, has been taken into account for the first
time.

\vspace{10mm}

\vfill

$^*$ e-mail: tanya@lukash.asc.rssi.ru

$^{**}$ e-mail: aal@hea.iki.rssi.ru

\clearpage

\section*{INTRODUCTION}

\vskip -5pt

Detection of the cosmic microwave background anisotropy (WMAP)and
investigation of the spatial distribution of galaxies and their
clusters (SDSS) (Tegmark et al. 2004) have allowed the most probable
values of the most important cosmological parameters, such as the
baryon density, the total matter density, and the cosmological
constant, to be obtained. Analysis of the observational data has
revealed that the visible matter accounts for only a few percent of
the total mass of the Universe and the observation of dark matter and
the determination of its nature are among the key problems in
cosmology. Analysis of the rotation curves for spiral galaxies from
Rubin et al. (1980) indicated that the mass of the galaxies required
to explain these curves is a factor of 10 larger than the total mass
of all the known spiral galaxy components (stars, gas, dust). In
addition, Ostriker et al. (1974) showed that the galactic disks are
unstable without a massive halo whose composition is unknown. These
conclusions are also valid for our Galaxy; it is believed that its
disk does not contain any signi cant amount of dark matter, while more
than 90\% of the halo matter is unobservable (Kuijken and Gilmore
1991; Roulet and Mollerach 1997). This dark halo may consist of both
baryonic and nonbaryonic matter. In the former case, massive compact
halo objects (MACHOs), such as brown dwarfs, primordial black holes,
white dwarfs, Jupiter-like planets, etc., may constitute the dark halo
mass (Carr 1994). In the latter case, the nonbaryonic matter may be
represented by weakly interacting massive particles (WIMPs) that are
also capable of clumping into compact dark matter objects (Berezinskiy
et al. 2003).

The problem of dark matter detection also exists on smaller scales
than the scale of the Universe. As was shown by Heggie and Hut (1996),
low-mass stars and white dwarfs, which are unobservable because of
their low luminosity, constitute half of the mass of globular star
clusters. In contrast to the population of light objects, heavy stars
tend to sink to the cluster core as a result of mass stratification.
Consequently, one might expect the dark component of a cluster to be
detected on its periphery. The presence of a significant amount of dark
matter forces the researchers to search for methods of its
detection.

One of these methods of searching for dark mass in the Galaxy is to use the
characteristic variability of the light curves for certain stars in the
Large and Small Magellanic Clouds (Paczynski 1986), i.e., to search for
microlensing events. The necessity of monitoring millions of stars and the
fact that the detection probability of microlensing events is low are the
main problems of this method. The possibility of using extragalactic optical
pulsars in searching for dark galactic objects was also discussed (Schneider
1990). When the signal from a pulsar passes near a massive compact object,
its ux is magnified by gravitational lensing. However, apart from flux
magnification, a time delay of the signal will also be observed in this case
(Krauss and Small 1991). Thus, in addition to gravitational lensing, another
general relativity effect, namely, the relativistic time delay of the signal
from a pulsar in the gravitational field of a massive body can be used to
detect both baryonic and nonbaryonic dark matter objects, which was
suggested by Larchenkova and Doroshenko (1995). The relativistic travel time
delay of an electromagnetic signal in a static, spherically symmetric
gravitational field of a point mass is called the Shapiro effect. This, in
turn, allowed the parameters of the gravitating body to be
determined. Subsequently, Wex et al. (1996) considered the possibility of
using pulsars behind the Galactic center as a test for studying the mass
distribution in the Galactic bulge; Fargion and Conversano (1997) described
the gravitational Shapiro phase shift on pulsar's period for the detection
of the dark matter. The Shapiro effect that results from the passage of a
star close to the line of sight to the pulsar was suggested as a possible
cause of the pulsar glitches (Sazhin 1986).

The detection of a significant number of pulsars in the globular
clusters 47 Tuc, Terzan 5, and M15 (Freire 2006) makes it
interesting to discuss the question of using them to study the dark
matter both in the globular clusters themselves and in the Galaxy
along the line of sight toward these clusters. The lowfrequency
pulsar timing noise in globular star clusters produced by the
Shapiro effect and attributable to random passages of cluster stars
near the line of sight to the pulsar was considered by Larchenkova
and Kopeikin (2006). These authors also obtained spectral
characteristics of this noise process, in particular, the slope of
the power spectrum. The latter will allow the noise due to the
stochastic Shapiro effect to be distinguished from the low-frequency
timing noise of a different nature in the observational data.
However, the detection probability of individual relativistic pulsar
signal time delay events in globular clusters caused by the passages
of both visible and dark objects of the globular cluster itself and
Galactic compact objects close to the line of sight remains unclear.

In this paper, we investigate the possibility of observing and
detecting single relativistic signal time delay events for pulsars
in globular star clusters. In the next section, we give a brief
overview of the signal magnification and angular splitting results
for the lensing of a point mass and obtain, in the gravitational
lens approximation, numerical estimates of the pulsar pulse time
delay in the gravitational field of this mass. Next, we provide
brief information about the pulsars in globular star clusters and
summarize the calculations of the probability of observing
relativistic signal time delay events for pulsars in the clusters 47
Tuc, Terzan 5, and M15 for large impact parameters compared to the
Einstein-Chwolson radius both on objects of the clusters themselves
and on Galactic objects. In the Sections "Number of Stars" and "The
Event Rate", we estimate the number of objects located close to the
line of sight to the source and producing the characteristic PAT
modulation and the rates of these events, respectively. The results
obtained are discussed in Conclusion.\\


\section*{ANGULAR SPLITTING, MAGNIFICATION, AND TIME DELAY OF THE PULSAR
SIGNALS}

Let us consider the classical model of gravitational lensing for a
point lens (a deflecting body) with mass $M$. The geometry under
consideration is presented in Fig. 1. The geometrical position of
the pulsar (PSR) in the sky is specified by the angle $\theta_s$,
while the positions of its images are specified by the angles
$\theta_{+}$ and $\theta_{-}$, where the '$+$' and '$-$' signs
correspond to the first ($+$) and second ($-$) images, respectively,
$d$ is the impact parameter of the undeflected light ray. The
Einstein-Chwolson radius is defined by the formula (Einstein 1965)
\begin{equation}
R_E=(4GMD_{ds}D_d/c^2 D_s)^{1/2},
\end{equation}
\noindent where $c$ is the speed of light in a vacuum, $G$ is the
gravitational constant, $D_s$ is the distance from the pulsar (PSR)
to the observer (O), $D_{ds}$ and $D_d$ are the distances from the
pulsar to the massive body ($M$) and from this body to the observer,
respectively. Three effects are well known for the classical model
under consideration: (1) pulsar image splitting in the plane of the
gravitating body, (2) flux magnification, and (3) sign al time
delay. For microlensing (lensing by stars), the angular image
splitting is much smaller than $10^{-3}$ arcsec and it cannot yet be
resolved with currently available instruments. The flux
magnification for the lensed images is given by the formula (Refsdal
1964):
\begin{equation}
 \mu_{+,-}=\frac{1}{4}\left[\frac{f}{(f^2+4)^{1/2}}+
 \frac{(f^2+4)^{1/2}}{f} \pm 2\right]\ ,
\end{equation}
\noindent where $f$ is the dimensionless impact parameter, $f =
d/R_E$. It is easy to see from Eq. (2) that significant pulsar flux
magnification takes place only if the impact parameter d falls into
the Einstein–Chwolson ring, i.e., $d < R_E$. When $f\gg 1$, the
contribution from the second ($-$) image to the total brightness is
small and this image is too faint to be observable.

Strict pulsar pulse periodicity allows the PAT variations to be used
to detect the massive bodies located close to the signal propagation
path from the pulsar to the observer. The coordinate travel time
delay of the light ray along the trajectory deflected by a
gravitating body relative to the undeflected trajectory is given by
the formula (Schneider et al. 1992)
\begin{equation}
c \Delta t = \frac{D_d D_{ds}}{2D_s} {\left[\hat \alpha ({\bf
\xi})\right]}^2 - \hat \psi({\bf \xi}) + const ,
\end{equation}
\noindent where $\hat \psi$ is the deflecting potential, {\bf $ \hat
\alpha$} is the deflection angle, and {\bf $\xi$} specifies the ray
location in the plane of the gravitating body. As we see from this
formula, the time delay consists of three components: the
geometrical time delay (the first component), the relativistic time
delay (the second component), and the constant calculated by
Kopeikin and Schafer (1999). For a spherically symmetric
Schwarzschild lens, Eq. (3) can be rewritten as
\begin{equation}
\Delta t = \frac{2GM}{c^3} \left[ \frac{4}{(\sqrt{f^2+4}\pm f)^2} -
\ln(\sqrt{f^2+4} \pm f)^2 \right]+const.
\end{equation}
\noindent

Since we are interested in the case of $f\gg 1$, for which, as was
mentioned above, only one pulsar image (signal) is observed, it
follows from Eq. (4) that the geometrical time delay of the first
ray ($+$) may be neglected compared to the relativistic time delay.
Since the pulsar and the massive body are moving, it is convenient
to estimate the Shapiro effects in the gravitational lens
approximation in terms of the pulsar velocity $V_P$ projected onto
the plane of the sky relative to the massive body ($M$) and the time
of the closest approach of this body to the line of sight ($T_0$) .
The relativistic signal time delay can then be written as
(Larchenkova and Doroshenko 1995)
\begin{equation} \Delta t = - \frac{2GM}{c^3}\cdot
             \ln(1+{\beta_0}^2\cdot (t-T_0)^2) \ ,
\end{equation}
\noindent where we use the designation ${\beta_0}\equiv V_P/d$. Let
the pulsar PAT observation begin at time $t = t_0$. The maximum
relativistic time delay occurs at the $t = T_0$ that corresponds to
the minimum impact parameter, which is a function of time. Suppose
that the minimum measurable Shapiro delay is 0.5 $\mu$s, which
corresponds to the current status of millisecond pulsar timing. For
our case of large impact parameters, the observing time $(t_0-T_0)$
it takes for the specified relativistic time delay to be recorded is
given in Table 1 for various masses of the deflecting body, impact
parameters, and pulsar velocities relative to the gravitating body.
Note that the velocities used in the table are close to the typical
velocities of the objects in globular clusters and the Galaxy,
respectively. The corresponding maximum Shapiro delays that can be
observed in the characteristic time $t_0-T_0 = 5$ years are given in
Table 2 for various gravitating masses, impact parameters, and
relative pulsar velocities.
\begin{table}[t]
\begin{center}
\renewcommand{\arraystretch}{1.1}
\caption{Observing time ($t_0-T_0$) it takes for the Shapiro effect
delay $\Delta t_{min}=0.5 \mu$s to be recorded for various masses of
the gravitating body $M$ and impact parameters $d$.}
\bigskip
\begin{tabular}{|c|c|c||c|c|}
\hline {} & \multicolumn{2}{|c||}{$V_P$=30 km s$^{-1}$} &
 \multicolumn{2}{|c|}{$V_P$=200 km s$^{-1}$}\\
\hline
M & d=10 AU & d=100 AU & d=10 AU & d=100 AU\\
\hline
$1 M_{\odot}$ & 128 days & 3.5 years & 20 days & 197 days\\
\hline
$0.5 M_{\odot}$ & 179 days & 4.9 years & 28 days & 281 days\\
\hline
$0.1 M_{\odot}$ & 1 year & 10 years & 69 days & 1.9 days\\
\hline
\end{tabular}

\end{center}
\end{table}
\begin{table}[h]
\begin{center}
\renewcommand{\arraystretch}{1.1}
\caption{Maximum Shapiro delay observed in 5 years for variousmasses
of the gravitating body $M$ and impact parameters $d$.}
\bigskip
\begin{tabular}{|c|c|c||c|c|}
\hline \multicolumn{3}{|c||}{$V_P$=30 km s$^{-1}$} &
\multicolumn{2}{|c|}
{$V_P$=200 km s$^{-1}$}\\
\hline
M & d=10 AU & d=100 AU & d=10 AU & d=100 AU\\
\hline
$10 M_{\odot}$ & 237 $\mu$s & 10 $\mu$s & 541 $\mu$s & 129 $\mu$s\\
\hline
$1 M_{\odot}$ & 23.7 $\mu$s & 1 $\mu$s & 54.1 $\mu$s & 12.9 $\mu$s\\
\hline
$0.5 M_{\odot}$ & 11.9 $\mu$s & 0.5 $\mu$s & 27 $\mu$s & 6.5 $\mu$s\\
\hline
$0.1 M_{\odot}$ & 2.4 $\mu$s & 0.1 $\mu$s & 5.4 $\mu$s & 1.3 $\mu$s\\
\hline
\end{tabular}

\end{center}
\end{table}
Let us introduce a new quantity, $d_{max}$, the maximum value of the
impact parameter $d$ at which the relativistic time delay is still
observable in the pulsar PAT residuals. As follows from Eq. (5),
$d_{max}$ is a function of the mass of the deflecting body $M$, the
relative pulsar velocity $V_P$, and the minimum observable Shapiro
delay $\delta t_{min}$. Consider two possibilities: the massive body
is located in a globular star cluster (1) and in the Galaxy along
the line of sight it (2). Let the observing time $(t_0-T_0) = 5$
years, $\delta t_{min}= 0.5 mu$s, $V_P = 30$ km s$^{-1}$ in the
first case and $V_P = 200$ km s$^{-1}$ in the second case. We use
the designations $d_{max1}$ and $d_{max2}$ corresponding to these
two cases. The derived expressions for $d_{max1}$ and $d_{max2}$
depend only on the mass of the deflecting body and can be well
approximated in a wide range of masses by a function of the form
$d\simeq A m^{\alpha}$ with the following parameters:
\begin{equation}
d_{max1}[a.e.] \simeq \left\{
\begin{array}{l}
138 \cdot m^{0.45},  0.08 \le m < 0.5\\
142 \cdot m^{0.5}, ~~  0.5 \le m < 10
\end{array}\right.\nonumber
\end{equation}
\begin{equation}
d_{max2}[a.e.] \simeq \left\{
\begin{array}{l}
922 \cdot m^{0.45},  0.08 \le m < 0.5\\
948 \cdot m^{0.5}, ~~  0.5 \le m < 10
\end{array}\right.
\end{equation}
\noindent where $m$ is the mass of the deflecting body in solar
masses. Since the number of stars decreases rapidly with increasing
stellar mass (see the mass function in the Section "The Probability
of a Single Relativistic Delay Event"), we limited the range of
masses under consideration from above by $10 M_{\odot}$.\\

\section*{PULSARS IN GLOBULAR CLUSTERS}

Twenty four globular star clusters containing 129 pulsars are known
to date in our Galaxy (Freire 2006). The overwhelming majority of
these pulsars have spin periods shorter than 25 ms and more than two
thirds of them are members of binary systems. Respectively, 8
(Anderson 1992), 22 (Lorimer et al. 2003), and 32 (Ransom et al.
2005) pulsars have been discovered in the globular clusters Ì15 (NGC
7078), 47 Tuc (NGC 104), and Terzan 5. All of the pulsars in Ì15,
except for B2127+11C that is a member of a binary system, are
concentrated in the cluster core (Anderson 1992). In addition, all
of the pulsars in 47 Tuc are within 1.2 arcmin of the cluster center
(i.e., within three cluster core radii)and have spin periods up to 8
ms (Lorimer et al. 2003). The globular star clusters in which
pulsars have been discovered have a dense core and, in most cases, a
large total mass, which manifests itself in high star escape
velocities from the cluster. For example, the escape velocity
$v_{esc}$ is 58 km s$^{-1}$ for 47 Tuc and 55 km s$^{-1}$ for M15
(Webbink 1985; Gebhardt al.1997). The mean proper motion of the
millisecond pulsars is $87\pm13$ km s$^{-1}$, which is a factor of 3
smaller than that of the normal pulsars (Hobbs et al. 2005). The
proper motions were measured for some of the pulsars in globular
clusters. For example, the mean 2D pulsar speed for 11 pulsars in 47
Tuc corrected for the proper motion of the cluster itself is
$25\pm5$ km s$^{-1}$ (Hobbs et al. 2005), in good agreement with the
escape velocity from the cluster (58 km s$^{-1}$). Based on these
measurements, we used the transverse pulsar velocity of 30 km
s$^{-1}$ in our estimates.\\
\begin{table}[h!tbp]
\begin{center}
\renewcommand{\arraystretch}{1.1}
\caption{Basic parameters of the globular clusters M15, 47 Tuc, and
Terzan 5. }\label{tabl_clst}
\bigskip
\begin{tabular}{|r|c|c|c|}
\hline {} & M15 & 47 Tuc & Terzan5 \\
\hline $l$; $b$ ($^\circ$) &
65.01; -27.31 & 305.9; -44.89 & 3.84; 1.67 \\
$D_c$ (kpc) & 10.2 & 4.1 & 10.3 \\
$r_c$ (pc) & 0.07 & 0.52 & 0.40 \\
$\sigma$ (êì/ñ) & 11.6 & 11 & 10.6 \\
$\rho_{0,gc}$ ($M_{\odot}\text{pc}^{-3}$)
 & $2\times 10^6$ & $6\times 10^4$ &
$5\times 10^5$ \\
$r_t$ (pc) & 60.8 & 60.3 & 21.9 \\
$D_{GC}$ (kpc) & 11.2 & 9.1 & 1.0 \\
\hline
\end{tabular}

\end{center}
\end{table}

\section*{THE PROBABILITY OF A SINGLE RELATIVISTIC DELAY EVENT}

If the optical depth of the effect is so small that the event occurs
only once, then the differential probability of the event can be
written as $dP=n(x)\pi (fR_E)^2dx$, where $n(x)$ is the number
density of deflecting bodies with mass $M$ (Krauss and Small 1991).
The probability of a relativistic delay on all of the massive
objects between the pulsar and the observer can then be written as:
\begin{equation}
\ P = \int_0^{D_s} \frac{4\pi G}{c^2} f^2 \rho (x)
     \frac{(D_s -x) x}{D_s}\,dx\,
\end{equation}
\noindent where $\rho (x)=n(x)M$. Note that this probability depends
on the total mass of all deflecting bodies and does not depend on
the mass of an individual object. Let us now calculate the
probabilities of a signal delay event for two cases: (1) the pulsar
and the massive body are located in a globular star cluster; (2) the
pulsar is still located in the cluster, while the massive body is
located in the Galaxy along the line of sight to the cluster. The
latter case is also applicable to the Galactic pulsars that do not
belong to globular star clusters.\\

\subsection*{The Pulsar and the Massive Body in the Globular
Cluster}

Let us consider the globular cluster as a selfgravitating isothermal
sphere of identical stars (see, e.g., the calculations by Jetzer et
al. (1998)f or 47 Tuc). Basic parameters of the clusters M15, 47
Tuc, and Terzan 5 considered here are given in Table 3, where $l$
and $b$ are the Galactic coordinates, $D_c$ is the distance to the
cluster center, $r_c$ is the cluster core radius, $\sigma$ is the
stellar velocity dispersion of the cluster, $\rho_{0,\text{gc}}$ is
the central density of the cluster core, $r_t$ is the tidal radius,
and $D_{GC}$ is the distance from the Galactic center (GC) to the
cluster (Gebhardt et al. 1997;Webbink 1985; Jetzer et al. 1998). To
estimate the probability of signal time delay events, we will use a
simple cluster model in which the mass density as a function of the
cluster radius $R$ is given by
\begin{equation}
\rho_{\text{gc}} (R) =
\frac{\rho_{0,\text{gc}}}{1+\left(\frac{R}{r_c}\right)^2}.
\end{equation}
The event probability on cluster objects will then be written as the
following integral (see Eq. 7)
\begin{equation}
\ P = \int_{x_t}^{D_s} \frac{4\pi G}{c^2} f^2 \rho_{\text{gc}} (x)
     \frac{(D_s-x) x}{D_s}\,dx\,
\end{equation}
\noindent where the integration terminates at the cluster boundary
$x_t=D_c - \sqrt{{r_t}^2 - {D_c}^2 {\sin{\beta}}^2}$. Here, $\beta$
is the angle between the directions of the cluster center and the
source (for small angles considered in our problem,
$\sin{\beta}\simeq\beta$).

The ratio of the probability of a relativistic delay event to the
square of the dimensionless impact parameter $f$, $P/f^2$, as a
function of the source position in the globular cluster along the
observer's line of sight to the cluster center (i.e., at $\beta=0$)
depending on the radial distance to the cluster center is indicated
by different curves in Fig. 2 for each of the clusters under
consideration. The characteristic dependence of $P/f^2$ on the
angular distance between the directions of the source and the
cluster center, i.e., in a direction perpendicular to the line of
sight, is shown in Fig. 3 for a pulsar at the distance of the center
of the globular cluster Terzan 5. For clarity, Fig. 4 shows the 3D
distribution of the ratios of the probability of signal time delay
events to the square of the dimensionless impact parameter $f$ as a
function of the angular and radial distances to the cluster center
for Terzan 5.

If the pulsar lies at the globular cluster center, i.e., $\beta=0$
and $D_s = D_c$, then $P/f^2$ is $4.65\times10^{-8}$ for 47 Tuc,
$4.01\times10^{-8}$ for M15 and $1.92\times10^{-7}$ for Terzan 5. To
calculate the event probability itself, we must know the impact
parameter $f$, which, in turn, is a function of the mass of the
deflecting body and $d_{max1}$. For the subsequent estimates, we
will make the following significant simplifying assumptions: the
mass function for globular cluster stars does not depend on the
positions of the stars in the cluster and all of the cluster objects
deflecting the electromagnetic signal from the pulsar have the same
mass $0.3 M_{\odot}$ or $0.6 M_{\odot}$. These values are variously
estimated (see, e.g., Chabrier 2003 and references therein) to be
close to the mean masses of the globular cluster stars and, hence,
using them in our calculations seems quite justifiable and close to
reality. The latter assumption is also confirmed, for example, by
the fact that the population of low-mass white dwarfs in 47 Tuc
accounts $\sim50$\% of the total cluster mass (Heggie and Hut 1996),
while the white dwarfs in M15 constitute the vast bulk (in mass)of
the entire cluster population (about 85\%; Gebhardt et al. 1997).

Let us consider the dimensionless impact parameter $f$ ($f\gg 1$)as
a function of the distance between the deflecting body with a mass
of $0.3 M_{\odot}$ or $0.6 M_{\odot}$ and pulsar that are both
located in the globular cluster for two impact parameters, $d = 10$
and $d = 80$ AU. In Fig. 5, these dependences are shown in a wide
range of distances covering the linear sizes of all three clusters
under consideration ($0-120$ pc). Assuming the distance between the
massive body and the pulsar to be half the tidal cluster radius, we
will obtain, as an example, the following estimates for the
probability of a pulsar signal time delay event for the deflecting
body with a mass of $0.6 M_{\odot}$ and the impact parameter $d =
80$ AU: $P\sim 2\times 10^{-3}$ for 47 Tuc, $P\sim 1.7\times
10^{-3}$ for M15, and $P\sim 2.5 \times 10^{-2}$ for Terzan 5 (the
pulsar is located at the cluster center). For the deflecting body
with a mass of $0.3 M_{\odot}$, our estimates increase by a factor
of $\sim1.4$. Note also that since the matter in the globular
clusters concentrate strongly to the cluster center, the
characteristic distance between the deflecting body and the pulsar
is much less than half the tidal radius, which also causes the
probability of such events to increase significantly (see Fig. 5).\\

\pagebreak

\subsection*{The Pulsar in the Cluster and the Massive Body
outside the Cluster}

Let us now consider the case where the pulsar is in the globular
cluster, while the deflecting body is outside the cluster. As was
noted above, the calculations and estimates given below also remain
valid for the pulsars that do not belong to globular clusters.
Inaccurate knowledge of the parameters of the Galactic components
(disk, bulge, spheroid, halo, etc.)poses the main difficulty in
calculating the probability of relativistic delay events on Galactic
objects. A four-component model of the Galaxy consisting of a disk,
a central bulge, a spheroid, and a halo was first suggested by
Bahcall and Soneira (1980) and Bahcall (1986). In some way, this
model is "classical" and is still commonly used in many
calculations, including the calculations of weak lensing events
(see, e.g.,Wex et al. 1996). A large amount of observational data in
various wavelength ranges has appeared in recent years. These data
have made it possible to improve significantly our knowledge of the
Galactic structure and the parameters of the models that describe it
(see, e.g., Dwek et al. 1995; Dehnen and Binney 1998; Robin et al.
2003). Our subsequent calculations of the probability of
relativistic pulsar signal delay events, the number of massive
bodies near the line of sight, and the event rate for Galactic
objects are based on one of the most commonly used models suggested
by Dehnen and Binney (1998)(below referred to as DB). This model
includes a three-component disk that consists of thin and thick
disks and an interstellar medium, a flatted bulge, and a halo in the
following form:
\begin{eqnarray}
\rho_{\text{disk}}(r,z)=\rho_{0,\text{disk}}\exp\left[-\frac{R_m}{r}-\frac{r}{R_d}-\frac{|z|}{z_d}\right] \nonumber\\
\rho_{\text{bulge}}(r,z)=\rho_{0,\text{bulge}}\left(\frac{\sqrt{r^2+\frac{z^2}{q^2_b}}}{r_{0,b}}\right)^{-1.8}
           \exp\left[-\frac{r^2+\frac{z^2}{q^2_b}}{r_{t,b}}\right]\nonumber\\
\rho_{\text{halo}}(r,z)=\rho_{0,\text{halo}}\frac{r^2+\frac{z^2}{q^2_h}}{r_{0,h}^2}
\left(1+\frac{r^2+\frac{z^2}{q^2_h}}{r_{0,h}}\right)^{-4.207},
\end{eqnarray}
\noindent where $\rho_{\text{disk}}(r,z)$ describes each of the
three disk components with its own parameters
$\rho_{0,\text{disk}}$, $R_m$, $R_d$, $z_d$ and the total mass
$M_{\text{disk}}\simeq4.8\times10^{10} M_{\odot}$;
$\rho_{0,\text{bulge}}=0.7561 M_{\odot}/{\text{pc}^3}$,
$r_{0,b}=1~\text{kpc}$, $r_{t,b}=1.9~\text{kpc}$, $q_b=0.6$,
$\rho_{0,\text{halo}}=1.263 M_{\odot}/{\text{kpc}^3}$,
$r_{0,h}=1.09~\text{kpc}$, $q_h=0.8$.

For comparison, we also calculated the probabilities of relativistic
signal delay events for the "classical" model by Bahcall and Soneira
(below referred to as BS):
\begin{eqnarray}
\rho_{\text{disk}}(r,z)=\rho_{0,\text{disk}}\exp\left[\frac{R_{GC}-r}{3.5~\text{kpc}}-\frac{|z|}{125 \text{pc}}\right] \nonumber\\
\rho_{\text{bulge}}(r)=\rho_{0,\text{bulge}}\left(\frac{r}{1~\text{kpc}}\right)^{-1.8}
           \exp\left[-\left(\frac{r}{1~\text{kpc}}\right)^3\right]\nonumber\\
\rho_{\text{spher}}(r)=\rho_{0,\text{spher}}\frac{\exp\left[-b \left(\frac{r}{2.8~\text{kpc}}\right)^{1/4}\right]}{\left(\frac{r}{2.8~\text{êïê}}\right)^{7/8}}\nonumber\\
\rho_{\text{halo}}(r)=\rho_{0,\text{halo}}\frac{a^2+R^2_{GC}}{a^2+r^2},
\end{eqnarray}
\noindent where $R_{GC}$ is the solar Galactocentric distance
(assumed to be 8 kpc), $\rho_{0,\text{disk}}=0.04
M_{\odot}/{\text{ïê}^3}$ is the disk density in the solar
neighborhood, the bulge mass is $M_{\text{bulge}}\simeq10^{10}
M_{\odot}$, $\rho_{0,\text{spher}}\simeq1/500\rho_{0,\text{disk}}$,
$b=7.669$, $\rho_{0,\text{halo}}=0.01 M_{\odot}/{\text{pc}^3}$, $a$
is the core radius of the spherical dark matter halo. Its value is
believed to lie in the range from $\sim2$ to $\sim8$ kpc (Caldwell
and Ostriker 1981; Bahcall et al. 1983). Since this uncertainty has
a negligible effect on the estimate of the effect, we assume in our
estimates that $a\simeq2$ kpc.

It is easy to see from Eqs. (10)and (11) that the Galactic bulge
contributes significantly to the probability only for the pulsars
whose electromagnetic signal propagates in the immediate vicinity
($\lesssim 1-2$ kpc) of the Galactic center. It follows from Table.
3, which presents basic parameters of the globular clusters under
study, including the GC distance $D_{GC}$, that it is important to
take into account the influence of the bulge only for the pulsars in
Terzan 5, while the signal from the pulsars in the high-latitude
clusters M15 and 47 Tuc propagates toward the observer outside the
bulge. Note also that the contribution from the disk naturally
decreases with increasing Galactic latitude.

Figure 6(a,b) shows the ratios of the detection probability of
relativistic PAT time delay events to $f^2$ as a function of the
Galactic coordinates of the source for two distances between the
observer and the pulsar, 10.2 and 4.1 kpc (the DB model of the
Galaxy). These distances were chosen, because the latter matches the
distance to 47 Tuc, while the former roughly corresponds to the
distances to M15 and Terzan 5 (the positions of all three clusters
are indicated by the crosses and the asterisk). We see that the
detection probability of events for large distances increases
significantly as the Galactic center is approached due to an
enhanced concentration of objects in this region. Note, however,
that we excluded the innermost part of the bulge called the "Nuclear
Bulge" within $\sim30$ pc.

Since a supermassive black hole is present at the Galactic center,
the structure of this region is rather complex and cannot be
described by a single component (for more detail, see Launhardt et
al. 2002). Figure 6(c) shows the contributions from various Galactic
components to the total detection probability of events for a source
at a distance of 10.2 kpc and a Galactic latitude of $1.67$\deg\
(the latitude of the globular cluster Terzan 5). We see from the
figure that the contribution from the bulge is significant only at
$|l|\lesssim20$\deg.

For comparison, Fig. 7 shows the same dependences as those in Fig. 6
for the BS model of the Galaxy. On the whole, the picture is similar
to what has been obtained previously for the DB model. We only note
that the contribution from the spheroid is negligible compared to
the other components and the influence of the bulge is significant
in a narrower l range. The slightly higher probabilities for the BS
model (by $sim 10-40$\%) are attributable to a more compact bulge at
the same total mass $\sim10^{10} M_{\odot}$ as that in the DB model
and a more massive (by a factor of $sim1.5$) Galactic disk assumed
in the BS model than that in the DB model.

To conclude this section, note that the structure of the Galaxy and
its components is assumed to be axisymmetric in both models under
consideration. In contrast, observations (see, e.g., Dwek et al.
1995; Revnivtsev et al. 2006) suggest that there is a small
asymmetry ($sim10-15$\%)in the distribution of stars relative to the
Galactic center related to a rotated, elongated bulge in the central
regions of the Galaxy.\\

\section*{THE NUMBER OF STARS}

We can now calculate the expected number (N) of visible stars on the
pulsar signal propagation path to the observer that affect the PATs.
The mass functions for the stars of various Galactic components and
the globular cluster were taken from Chabrier and Mera (1997)and
Chabrier (2003) in the form the disk, bulge
\begin{equation}
\xi (\log{m})=\frac{1}{\log{M_\odot} \text{pc}^3} \left \{
\begin{array}{l}
0.158\exp\left[-\frac{\left( \log{\left(\frac{m}{0.079} \right)} \right)^2}{0.9577} \right],  0.08 \le m < 1 \\
4.4\times10^{-2} m^{-4.37},  1.0 \le m < 3.47\\
1.5\times10^{-2} m^{-3.53},  3.47 \le m < 10.0.
\end{array} \right.
\end{equation}
halo
\begin{equation}
\xi (m)=4\times10^{-3}\left(\frac{m}{0.1 M_\odot} \right)^{-1.7}
\frac{1}{M_\odot \text{pc}^3} , 0.01 \le m < 0.8
\end{equation}
globular cluster
\begin{equation}
\xi (\log{m})=\frac{1}{\log{M_\odot} \text{pc}^3} \left \{
\begin{array}{l}
3.6\times10^{-4}\exp\left[-\frac{\left( \log{\left(\frac{m}{0.33} \right)} \right)^2}{0.2312} \right],  m \le 0.9 \\
7.1\times10^{-5} m^{-1.3},  m > 0.9.
\end{array} \right.
\end{equation}
\noindent where the mass $m$ is in solar masses. It is easy to see
that the above formulas describe the stellar component of the
Galaxy, i.e., the objects with masses $>0.08 M_\odot$. However, a
large population of objects with masses $0.01<m\lesssim 0.08
M_\odot$, the so-called brown dwarfs, is currently believed to be
present in the Galaxy (particularly in its disk). Whereas the
contribution from such objects to the total mass of the Galaxy is
small, they can contribute noticeably to the estimate of the number
of stars affecting the PATs, since their concentration is comparable
to that of the normal stars. The mass function for brown dwarfs is
(Chabrier 2003)
\begin{equation}
\xi (\log{m})=0.158 m^{0} \frac{1}{\log{M_\odot} \text{ïê}^3} , 0.01
\le m < 0.08.
\end{equation}
\noindent The number of Galactic stars affecting significantly the
PATs can be written as
\begin{equation}
N = \frac{\int_{0}^{D_s} \int_{m_1}^{m_2} \pi \rho (x) \xi(m)
[d_{max2}(m)]^2 \,dx \,dm}{M_\odot \int_{m_1}^{m_2} m \xi(m) \,dm} ,
\end{equation}
\noindent where $\rho (x)$ the density of the sources in the Galaxy
(the DB model) along the line of sight to the pulsar, $(m1,m2)$ is
the mass range of objects under consideration, in our case, from
0.01 to $10 M_\odot$ if the brown dwarfs are taken into account.
Equation (16) is valid for calculating the number of stars affecting
the pulsar PATs and belonging to the globular cluster itself to
within the substitution of $d_{max1}$ for $d_{max2}$. Integrating
Eq. (16) in the entire range of masses and distances and taking into
account Eqs. (10) and (12)–(15), we obtain estimates for the
expected number of objects located close to the line of sight both
in the Galaxy (given the contribution from its various components)
and in the globular cluster (see Table 4).

\begin{table}
\begin{center}
\caption{}
\bigskip
\begin{tabular}{l|c|c|c|c|c}
\hline
       & disk  & bulge & halo  & Galaxy & cluster \\
\hline
M15    & 0.003 &  -    & 0.009 &  0.012    &  0.336    \\
47 Tuc & 0.002 &  -    & 0.004 &  0.006    &  0.075    \\
Terzan 5  & 0.273 &  0.166& 0.039 &  0.478    &  0.472    \\
\hline
\end{tabular}
\end{center}
\end{table}

We see that the expected number of Galactic objects near the line of
sight for the globular clusters M15 and 47 Tuc located far from the
Galactic center is considerably smaller than the number of objects
in the globular cluster itself. The two numbers are equal only for
the cluster Terzan 5 located immediately behind the Galactic center.
Note also the larger relative contribution from halo objects than
that from disk ones for the high-latitude clusters M15 and 47 Tuc.

It should be noted that one of the microlensing events detected by
the MACHO group is most likely caused by a lens located in the disk
(Alcock et al. 2001) and observed with the Hubble Space Telescope
(HST). The observations yielded estimates of the lens mass: this
mass either is $\sim0.04M_\odot$ or lies in the range $0.095-0.13
M_{\odot}$, i.e., the lens is either a brown dwarf or a low-mass
star.\\

\section*{THE EVENT RATE}

Undoubtedly, the detection probability of relativistic PAT time
delay events (or, in other words, the optical depth)is a very
important quantity. However, knowing the rate of such events is no
less important. By analogy with the rate of lensing events (Griest
1991), we will introduce the differential number of pulsar PAT time
delay events
\begin{equation}
dN_{ev} = N_s t_{obs} d\Gamma,
\end{equation}
\noindent where $N_s$ is the total number of sources (pulsars) for
an observing time $t_{obs}$, $d\Gamma$ is the differential event
rate,
\begin{equation}
d \Gamma = \frac{n({\bf{x}}) f({\bf{v}}_l) d^3 x d^3 v}{dt} ,
\end{equation}
\noindent where the numerator is the number of massive bodies in the
element of volume $d^3 x = dx dy dz$ and velocity space $d^3 v =
dv_x dv_y dv_z$ relative to the position $\bf{x}$ in the event
"tube", $n({\bf{x}})$ is the number density of massive objects, and
$f({\bf{v}}_l)$ is their velocity profile. The differential velocity
with which the gravitating masses with velocities ${{\bf v}_l}$
contribute to the event tube at the position $\bf{x}$ and the
cylindrical segment $l d \phi = d_{max} d \phi$ can be written as
\begin{equation}
d \Gamma = \frac{\rho (x)}{\left<M\right>} f({\bf{v}}_l) v^2_r cos
\theta d_{max} d\phi dv_x d v_r d \theta d x ,
\end{equation}
\noindent where $\left<M\right>$ is the mean mass of the deflecting
body. We also used the passage to cylindrical coordinates, $d^3 v =
v_r dv_x dv_r d\theta$, and the expression for $d_{max}$ derived
from Eq. (5),
\begin{equation}
d_{max}= v_r (t_0 - T_0) (e^ \frac{c^3 \Delta t_{min} }{2 G
\left<M\right>} - 1)^{-1/2} .
\end{equation}
\noindent As the velocity profile, we took a Maxwellian distribution
with a velocity dispersion $\sigma _l$,
\begin{equation}
f({\bf{v}}_l) d^3 v = \frac{1}{\pi ^{3/2} \sigma_l^3}
e^-{\frac{v_l^2}{\sigma_l^2}} d^3 v .
\end{equation}
Integration over the variables $d \phi, d v_x, d v_r$ and $d \theta$
yields
\begin{equation}
\Gamma = 4 \sigma^2_l (t_0 - T_0)\left<M\right>^{-1} (e^{ \frac{c^3
\Delta t_{min} }{2 G \left<M\right>}} - 1)^{-1/2} \int_0^{D_s} \rho
(x)\,dx
\end{equation}
As we see from Eq. (22), the event rate depends on the observing
time, the velocity dispersion of the objects, their mean mass, and
the matter distribution along the line of sight. Since these
parameters differ significantly, we calculated the event rate
$\Gamma$ separately for each Galactic component and the globular
cluster. The mean masses of the objects were calculated from the
above mass functions; their velocity dispersions $\sigma _l$ were
assumed to be 210, 100, 50, and 11 km s$^{-1}$ for the halo, bulge,
disk, and the cluster, respectively (see Zasov et al. 2004; Alcobe
and Cubarsi 2005; Vieira et al. 2006). The event rates are plotted
against the Galactic coordinates of the source in Figs. 8 and 9 for
two different heliocentric distances of the pulsar, 10.2 and 4.1 kpc
(we assumed that $\Delta t_{min} = 0.5 \mu$s and that the observing
time $t_0-T_0 = 5$ years). The positions of the globular clusters
M15, 47 Tuc, and Terzan 5 are indicated by the crosses and the
asterisk.

Table 5 gives the rates of relativistic PAT time delay events
(events per year)for pulsars in the clusters listed above in the
case of lensing by Galactic objects (the total value for all
components)and by objects of the cluster itself and the total number
of expected events $N_5$ for an observing time of 5 years.
\begin{table}
\begin{center}
\caption{}
\bigskip
\begin{tabular}{l|c|c|c}
\hline
       & Galaxy & Cluster & Number of events $N_5$\\
\hline
M15    & $1.15\times10^{-3}$    &  $3.4\times10^{-3}$  &  0.18 \\
47 Tuc & $5.44\times10^{-4}$    &  $7.6\times10^{-4}$  &  0.14 \\
Ter 5  & $1.05\times10^{-2}$    &  $4.8\times10^{-3}$  &  2.45 \\
\hline
\end{tabular}
\end{center}
\end{table}

Thus, by observing 22 pulsars in the globular cluster 47 Tuc, 8
pulsars in M15, and 32 pulsars in Terzan 5 for five years, one might
expect $\sim3$ relativistic PAT time delay events for pulsars to be
detected. The pulsars in the globular cluster Terzan 5 make a major
contribution to this number.

Note that the magnitude of the effect under consideration depends
strongly on $\Delta t_{min}$, which is determined by the accuracy of
the present-day PAT measurements and also depends on the intensity
of the specific pulsar. In particular, the expected number of
relativistic PAT time delay events for pulsars in the globular
cluster Terzan 5 for an observing time of 5 years decreases to
$N_5\sim0.75$ for $\Delta t_{min}= 2.5 \mu$s and to $N_5\sim0.4$ for
$\Delta t_{min}= 2.5 \mu$s.\\


\section*{CONCLUSION}

We considered the possibility of observing single relativistic
pulsar PAT time delay events caused by the passages of massive
objects close to the line of sight. Below, we briefly summarize and
discuss our most interesting and important results.

(1) We determined the probabilities of single relativistic PAT time
delay events for pulsars in three globular clusters, 47 Tuc, M15,
and Terzan 5, caused by the passages of massive bodies of the
cluster itself and Galactic objects on the path to the cluster near
the line of sight. Assuming the globular cluster mass density
distribution to be described by the model of an isothermal sphere
with a core, we found our ratios of the probability of delay events
on cluster objects to the dimensionless parameter f2 to be
comparable for the high-latitude globular clusters 47 Tuc and M15
and to be slightly higher for the more compact and dense cluster
Terzan 5.

(2) For the case where the massive body lies outside the cluster, we
calculated the probabilities of events for two models of the mass
distribution in the Galaxy: the "classical" Bahcall–Soneira model
(Bahcall and Soneira 1980; Bahcall 1986) and the more recent model
by Dehnen and Binney (1998). Our results are in good agreement with
one another; a certain excess of the probabilities for the BS model
is attributable to a more massive disk and a more compact bulge in
this model. The ratios of the probability of the events caused by
the passages of massive Galactic objects close to the line of sight
to the parameter $f^2$ for pulsars in the globular clusters 47 Tuc
and M15 are comparable to those for close passages of massive
objects in the clusters themselves. At the present accuracy of
measuring the expected effect from the pulsar PATs and an observing
time of about 5 years, the detection probability of such events
turns out to be low. The probability increases significantly only
for pulsars in the globular star cluster Terzan 5, since in this
case the line of sight passes through the dense bulge and disk
regions.

(3) We determined the number of objects near the line of sight
toward the pulsar that can produce the modulation of its PATs
characteristic of the effect under consideration located both in the
clusters under study and in the Galaxy on the path to the clusters.
The population of brown dwarfs in the Galactic disk, whose
concentration is comparable to that of the disk stars, has been
taken into account for the first time. As would be expected, the
number of massive objects near the line of sight is at a maximum for
the globular cluster Terzan 5.

(4) We calculated the rate of relativistic pulsar PAT time delay
events for the deflecting bodies in a globular cluster and the
Galaxy.

The accuracy in determining the main pulsar parameters, such as the
first, $\dot p$, and second, $\dot \dot p$, period derivatives, by
processing the timing data affects significantly the possibility of
detecting small relativistic effects. For most of the pulsars in
globular star clusters, only the first period derivative has been
measured well, $\sim 10^{-20}$ (the second period derivative has
been measured only for some of the sources, $\sim 10^{-31}$
s$^{-1}$; see, e.g., the catalog of pulsars and their parameters at
http://www.atnf.csiro.au). As regards the accuracy in determining
$\dot p$, it is presently $\sim0.1$\% (see, e.g., Freire et al.
2003) for the globular cluster 47 Tuc. The changes in $\dot p$ due
to the effect under consideration are
$\text{(several)}\times10^{-25}- 10^{-24}$ for the massive bodies
passing near the "event tube" boundary. In reality, this may prove
to be considerably larger because of the bodies passing closer to
the line of sight (depends as $\propto d^{-2}$ and may be comparable
to or even higher than the current accuracy in determining $\dot p$;
see also Wex et al. 1996).

All of the results obtained in this paper lead us to conclude that
the effect under consideration is small and difficult to measure at
the current level of observations for faint sources, which the
pulsars in globular clusters are; allowing for this effect will
become increasingly important with improving measurement accuracy.
However, this effect may turn out to be measurable for bright
pulsars even now at a pulsar PAT measurement accuracy of $\sim100$
ns in observing intervals of several years.

As was noted in the Introduction, invisible dark matter particles
are capable of clumping into compact objects with a mass of the
order of the Earth mass (Berezinskiy et al. 2003) and their
detection is of great interest to researchers. However, according to
our calculations, at the current level of accuracy, such objects
cannot be detected based on pulsar PAT measurements in a
"reasonable" observing time ($\sim10$ years) because of their low
mass (see Eq. (5)).

In conclusion, note that, apart from the Shapiro effect considered
here, the behavior of the pulsar PAT residuals can also be affected
by other effects, such as close passages of stars in globular
clusters and variations in the interstellar medium. The latter
effect depends on the frequency at which the observations are
performed and can be taken into account by performing observations
at several frequencies. Close passages of stars in globular clusters
can cause significant changes in both pulsar trajectories and
periods, $\Delta p/p \sim 10^{-8}$. However, the probability of such
events is low, because the mean distance between the cluster stars
is large, $\sim0.1$ pc (see, e.g., Rodin 2000).

\section*{ACKNOWLEDGMENTS}

We thank M.R. Gilfanov and M.G. Revnivtsev for a discussion of the
Galactic models used here and S.M. Kopeikin, B.V. Komberg, and V.N.
Lukash for helpful remarks and discussions. This work was supported
by a program of the Russian President (NSh-1100.2006.2) and the
Presidium of the Russian Academy of Sciences ("Origin and Evolution
of Stars and Galaxies"), grants of RFBR 07-02-01051, 07-02-00886 and
05-02-17465. A.A. Lutovinov is grateful to the Russian Science
Support Foundation. We wish to thank V.Astakhov for the help in
translating this paper in English.

\pagebreak

1. S. Alcobe and R. Cubarsi, Astron. Astrophys. 442, 929 (2005).

2. C. Alcock, R. Allsman, D. Alves, et al., Nature 414, 617 (2001).

3. S. J. Anderson, PhD Thesis (Caltech, 1992).

4. J. N. Bahcall, Ann. Rev. Astron. Astrophys. 24, 577 (1986).

5. J. N. Bahcall and R. M. Soneira, Astrophys. J., Suppl. Ser. 44,
73 (1980).

6. J. N. Bahcall, M. Schmidt, and R.M. Soneira, Astrophys. J. 265,
730 (1983).

7. V. Berezinskiy, V. Dokuchaev, and Y. Eroshenko, Phys. Rev. D 68,
103003 (2003).

8. J. A. R. Caldwell and J. P.Ostriker,Astrophys. J. 251, 61 (1981).

9. B. Carr, Ann. Rev. Astron. Astrophys. 32, 531 (1994).

10. G. Chabrier, Publ. Astron. Soc. Pac. 115, 763 (2003).

11. G. Chabrier and D.Mera, Astron. Astrophys. 328, 83 (1997).

12. W. Dehnen and J. Binney, Mon. Not. R. Astron. Soc. 294, 429
(1998).

13. E. Dwek, R. Arendt, M. Hauser, et al., Astrophys. J. 445, 716
(1995).

14. A. Einstein, Collection of Scientific Works in 4 Volumes
(Mir,Mocsow, 1965).

14. D. Fargion and R. Conversano, Mon. Not. R. Astron. Soc. 285, 225 (1997)

15. P. C. Freire, Pulsars in Globular Clusters, http://www.naic.edu/
pfreire/GCpsr.html (2006).

16. P. Freire, F. Camilo, M. Kramer, et al.,Mon. Not. R. Astron.
Soc. 340, 1359 (2003).

17. K. Gebhardt, C. Pryor, T.B.Williams, et al., Astron. J. 113,
1026 (1997).

18. K. Griest, Astrophys. J. 366, 412 (1991).

19. D. C. Heggie and P. Hut, in Proceedings of the IAU Symp. No. 174
(Kluwer, Dordrecht, 1996), p. 303.

20. G.Hobbs,D. Lorimer,A. Lyne, and M. Kramer,Mon. Not. R. Astron.
Soc. 360, 974 (2005).

21. Ph. Jetzer, M. Straessle, and U. Wandeler, Astron. Astrophys.
336, 411 (1998).

22. S. M. Kopeikin and G. Schafer, Phys. Rev. D 60, 124002 (1999).

23. L. M. Krauss and T. A. Small, Astrophys. J. 378, 22 (1991).

24. K. Kuijken and G. Gilmore, Astrophys. J. 367, L9 (1991).

25. T. I. Larchenkova and O. V. Doroshenko, Astron. Astrophys. 297,
607 (1995).

26. T. I. Larchenkova and S. M. Kopeikin, Pis'ma Astron. Zh. 32, 20
(2006)[Astron. Lett. 32, 18 (2006)].

27. R. Launhardt, R. Zylka, and P. Mezger, Astron. Astrophys. 384,
112 (2002).

28. D. R. Lorimer, F. Camilo, P. Freire, et al., Mon. Not. R.
Astron. Soc. 340, 1359 (2003).

29. J. Ostriker, P. J. E. Peebles, and A. Yahil, Astrophys. J. 193,
L1 (1974).

30. B. Paczynski, Astrophys. J. 301, 503 (1986).

31. S. M. Ransom, J. W. T. Hessels, I. H. Stairs, et al., Bull. Am.
Astron. Soc. 37, 1216 (2005); astroph/ 0501230.

32. S. Refsdal, Mon. Not. R. Astron. Soc. 128, 295 (1964).

33. M. Revnivtsev, S. Sazonov, M. Gilfanov, et al., Astron.
Astrophys. 452, 168 (2006).

34. A. C. Robin, C. Reyle, S. Derriere, and S. Picaud, Astron.
Astrophys. 409, 523 (2003).

35. A. Rodin, Cand. Sci. (Phys.-Math.) Disseration (Physical Lebedev
Institute, Academy of Science, Moscow, 2000).

36. E. Roulet and S. Mollerach, Phys. Rep. 279, 67 (1997).

37. V. C. Rubin,W. K. Ford, and N. Thonnard, Astrophys. J. 238, 471
(1980).

38. M. V. Sazhin, in Proceedings of the 11th International
Conference onGeneral Relativity and Gravitation, 1986, p. 519.

39. J. Schneider, New and Exotic Phenomena, Proceedings of XXV
Rencontre de Moriond, Ed. by O. Facker and J. Tran Tranh Van
(Frontieres, 1990), p. 301.

40. P. Schneider, J. Ehlers, and E. Falco, Gravitational Lenses XIV
560, 112 (1992).

41. I. I. Shapiro, Phys. Rev. Lett. 13, 789 (1964).

42. M. Tegmark, M. Strass, M. Blanton, et al., Phys. Rev. D 69,
103501 (2004).

43. K.Vieira, D.Dinescu,W. vanAltena, et al., Rev. Mex. Astron.
Astrofis. 25, 35 (2006).

44. R. F.Webbink, in Proceedings of the IAU Symp. No. 113: Dynamics
of Star Clusters, Ed. J. Goodman and P. Hut (Reidel, Dordrecht,
1985), p. 541.

45. N. Wex, J. Gil, and M. Sendyk, Astron. Astrophys. 311, 746
(1996).

46. A. V. Zasov, A. V. Khoperskov, and N. V. Tyurina, Pis'ma Astron.
Zh. 30, 653 (2004)[Astron. Lett. 30, 593 (2004)].

\pagebreak

\begin{figure}
\begin{center}
\includegraphics[width=12cm]{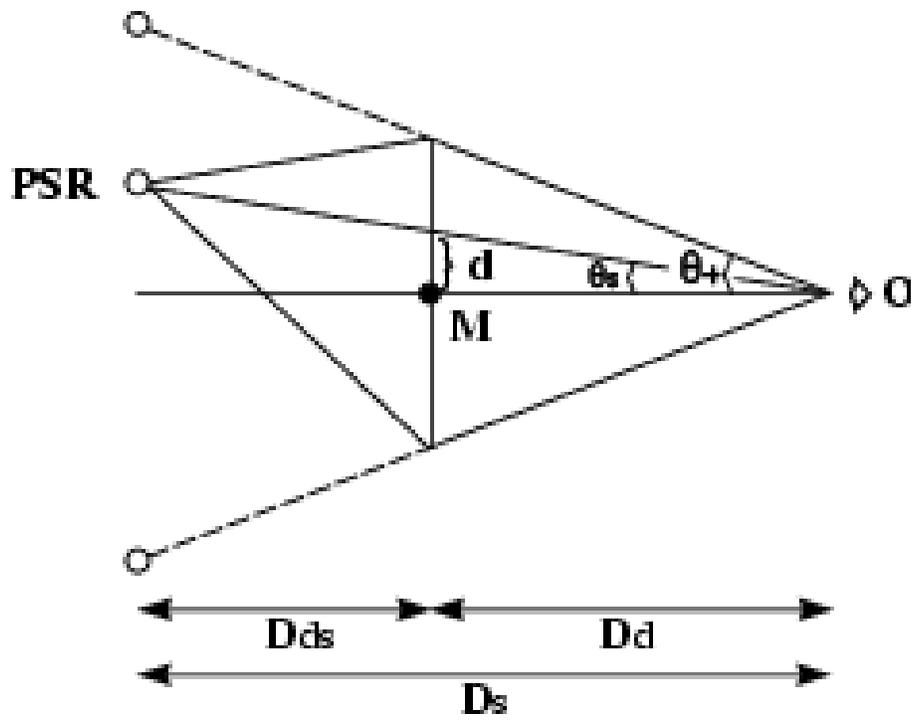}
 \caption{\rm Geometry of the problem under consideration:
O -— observer, PSR -— pulsar, Ì -— gravitating mass. For the
remaining notation, see the text.}
\end{center}
\end{figure}

\pagebreak

\begin{figure}
\begin{center}
\includegraphics[width=12cm,bb=35 150 550 700]{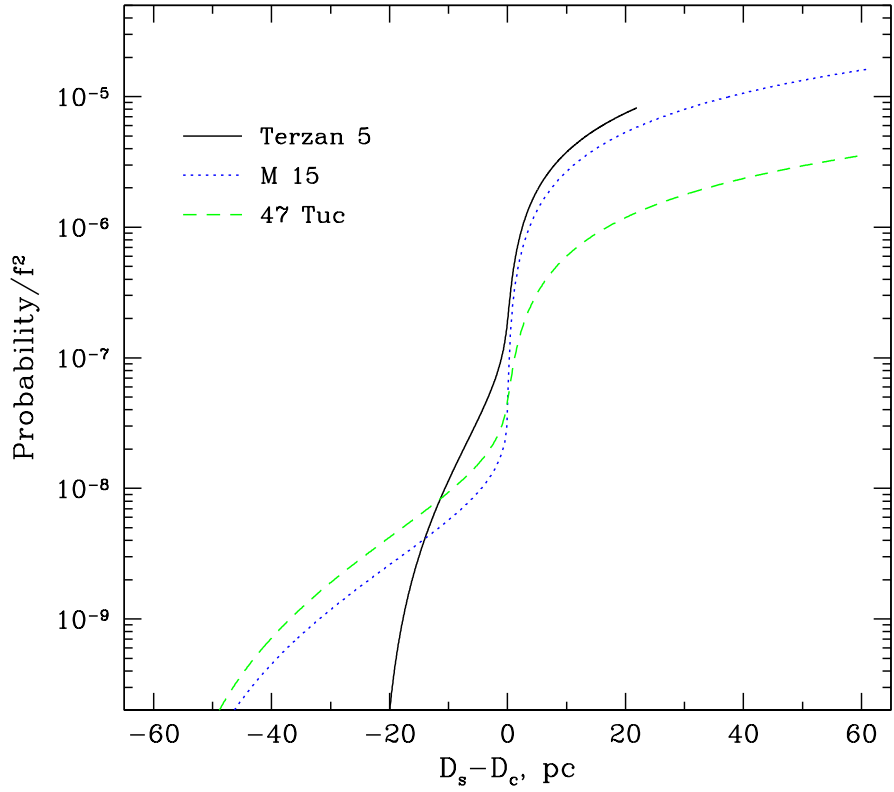}
 \caption{\rm Ratio of the detection probability of a time delay
event to $f^2$ as a function of pulsar position in the globular
clusters M15 (dotted line), 47 Tuc (dashed line), and Terzan 5
(solid line). All of the sources are located along the observer's
line of sight to the cluster center, but at different radial
distances from the cluster center.}
\end{center}
\end{figure}

\begin{figure}
\begin{center}
\includegraphics[width=12cm,bb=35 150 550 700 ]{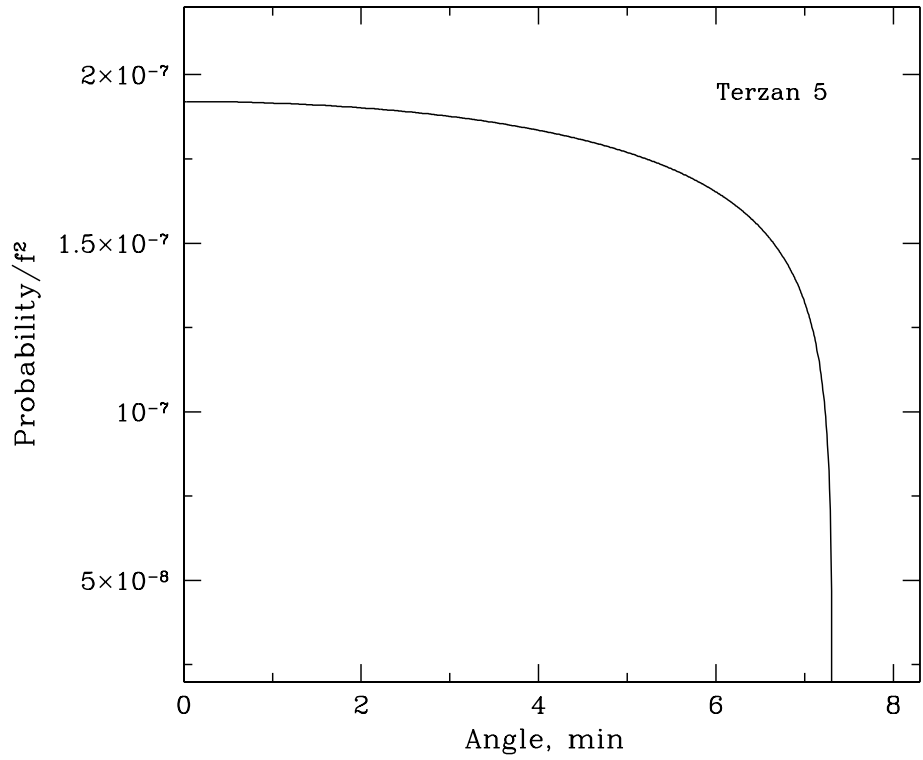}
\caption{\rm  Ratio of the detection probability of a time delay
event to $f^2$ as a function of pulsar position in the cluster
Terzan 5. All of the sources are located at the distance of the
cluster center, but at different angular distances from its center.
}
\end{center}
\end{figure}

\begin{figure}
\begin{center}
\includegraphics[width=15cm]{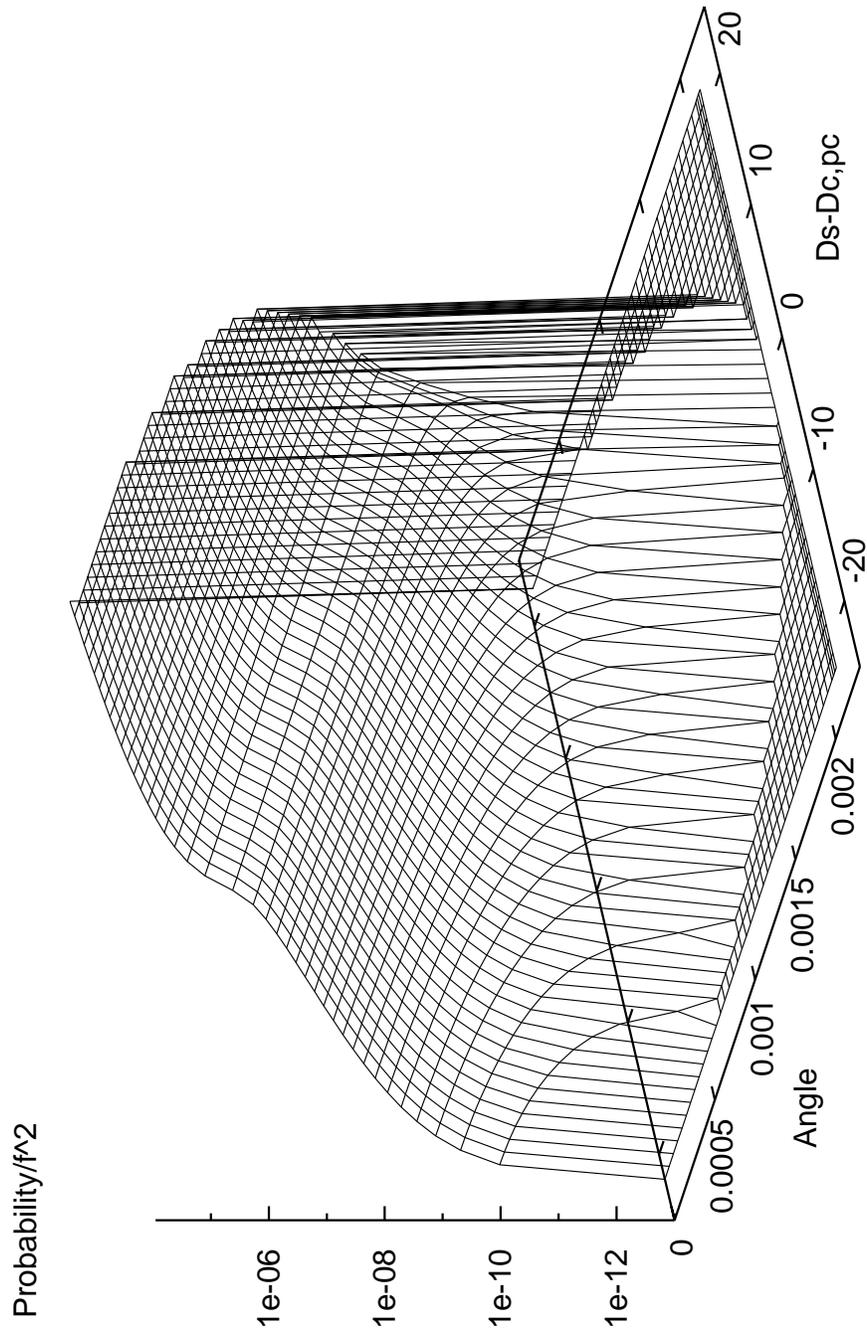}
\caption{\rm 3D distribution of the ratio of the detection
probability of a time delay event to $f^2$ for a source in the
globular cluster Terzan 5 as a function of its radial (pc) and
angular (rad) distances to the cluster center.  }
\end{center}
\end{figure}

\begin{figure}
\begin{center}
\includegraphics[width=12cm,bb=35 150 550 700]{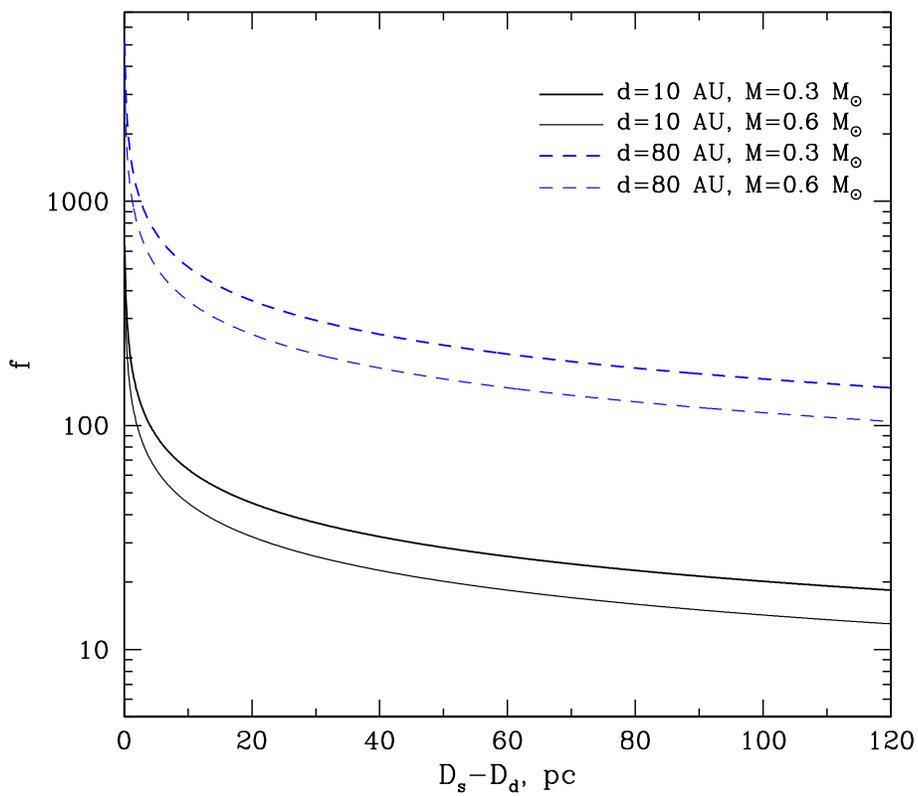}
\caption{\rm  Dimensionless impact parameter $f$ as a function of
the distance between the deflecting body with a mass of $0.3
M_\odot$ and $0.6 M_\odot$ and the pulsar for two different impact
parameters, $d = 10$ and $d = 80$ AU. }
\end{center}
\end{figure}

\begin{figure}
\begin{center}

\includegraphics[width=8.5cm,bb=35 265 545 680,clip]{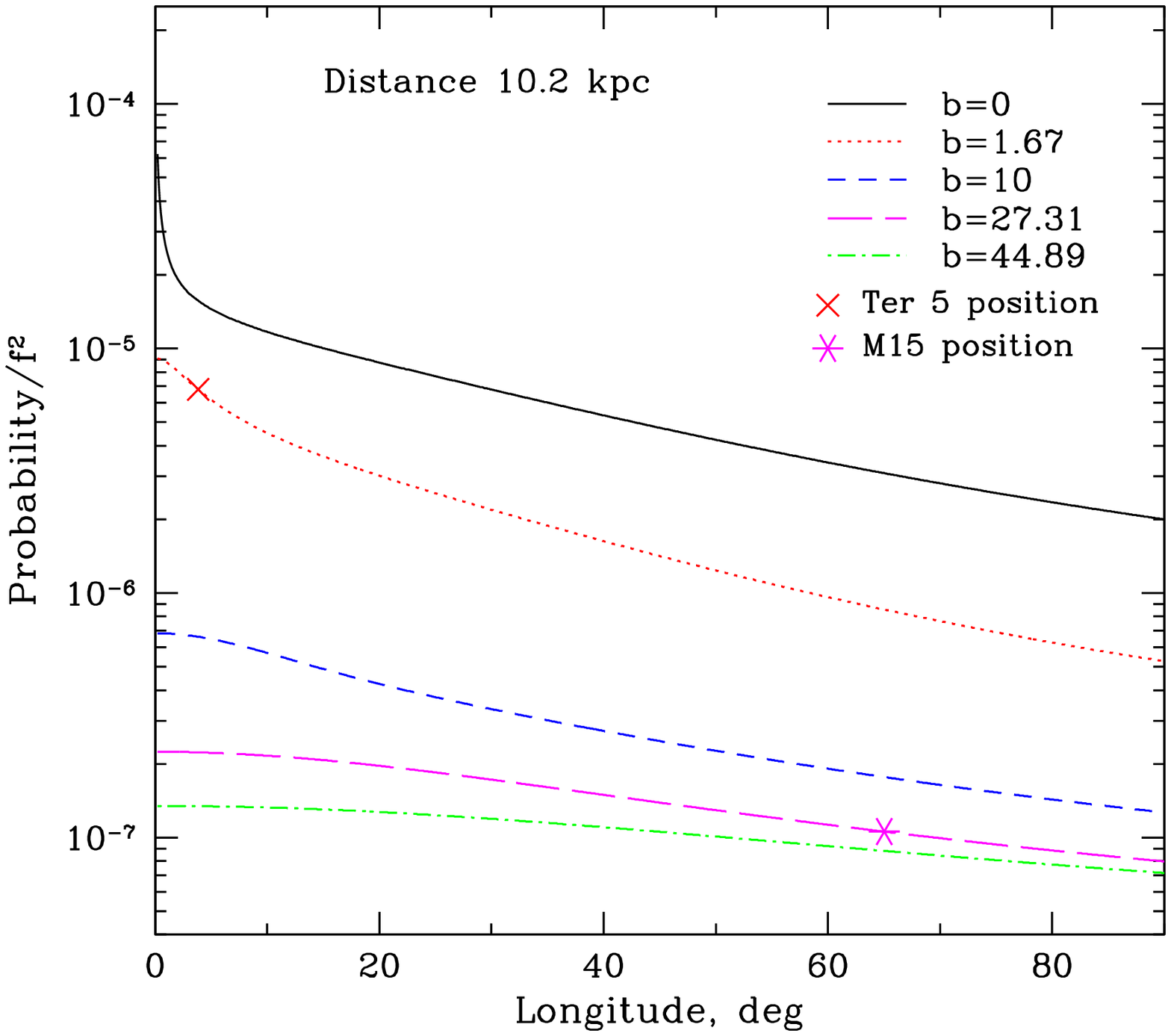}

\includegraphics[width=8.5cm,bb=35 265 545 680,clip]{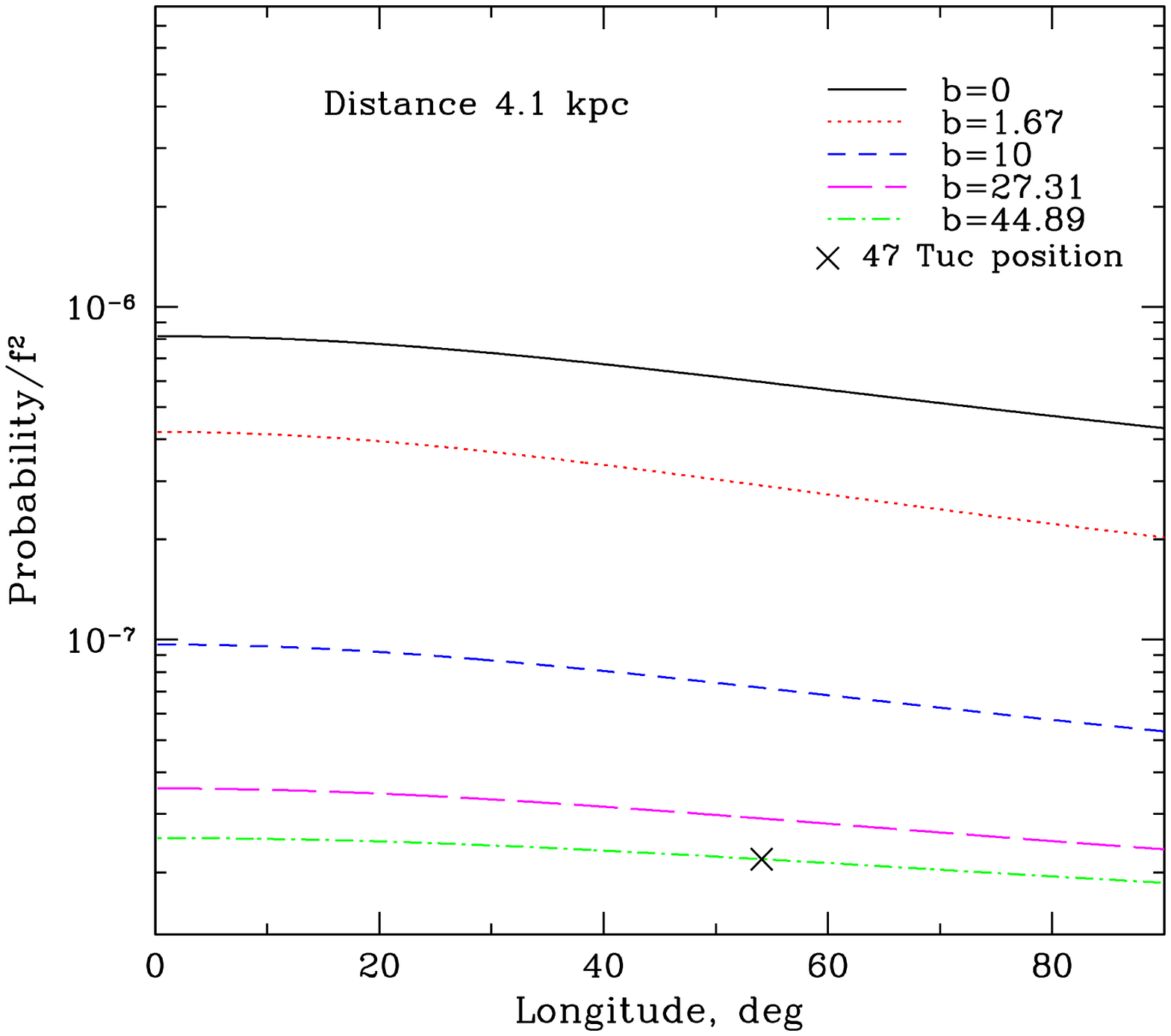}

\includegraphics[width=8.5cm,bb=35 215 545 680,clip]{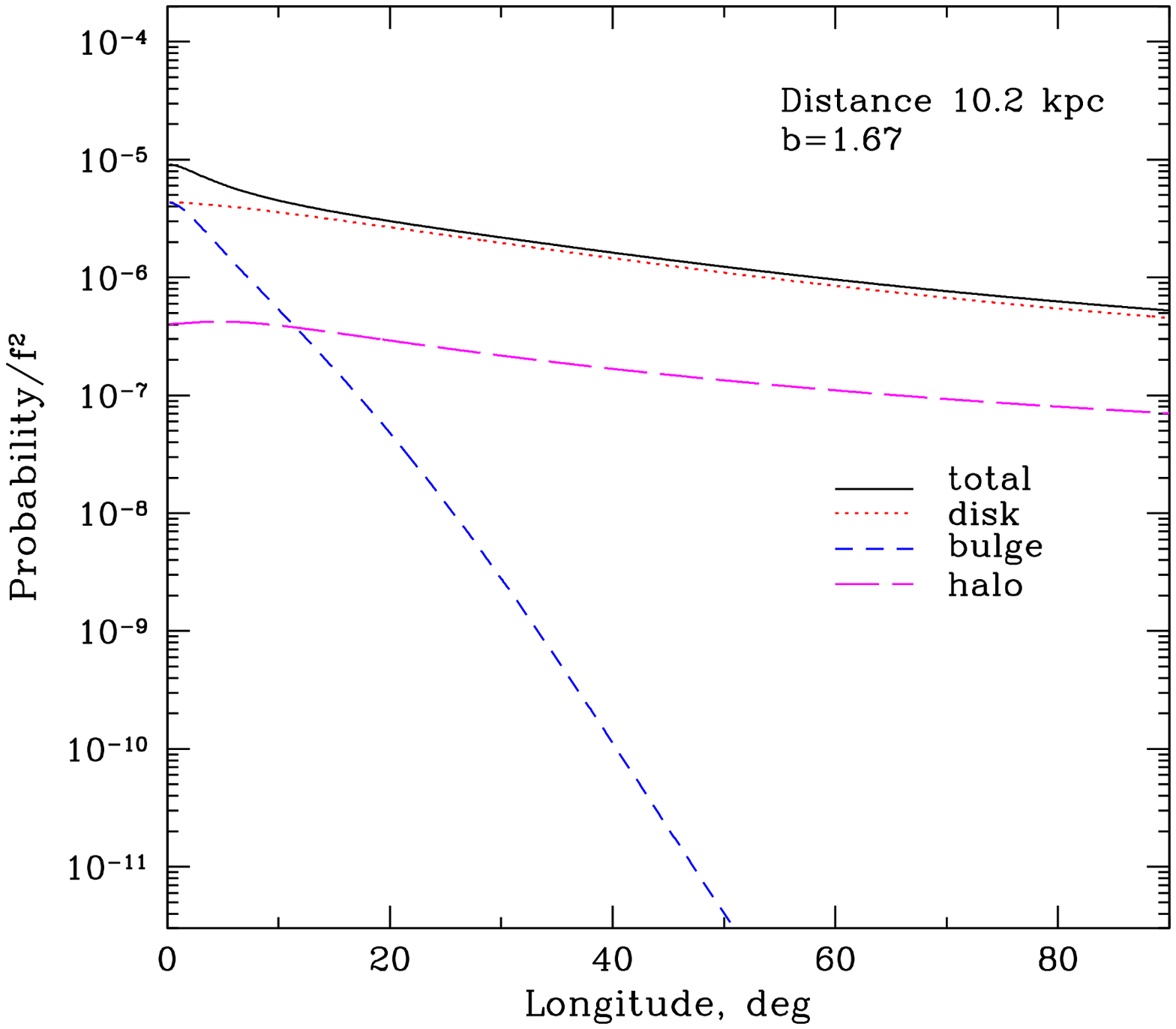}

\caption{\rm  Ratio of the detection probability of a time delay
event to $f^2$ as a function of Galactic longitude $l$ and latitude
$b$ of a source at a distance of (a) 10.2 and (b) 4.1 kpc. (c) The
relative contributions from various Galactic components to the total
lensing probability for a source at a latitude of 1.67\deg. The
positions of the globular clusters Terzan 5, M15, and 47 Tuc are
indicated by the crosses and the asterisk. The DB model of the
Galaxy is used. }

\end{center}
\end{figure}

\begin{figure*}
\begin{center}

\includegraphics[width=8.5cm,bb=35 265 545 680,clip]{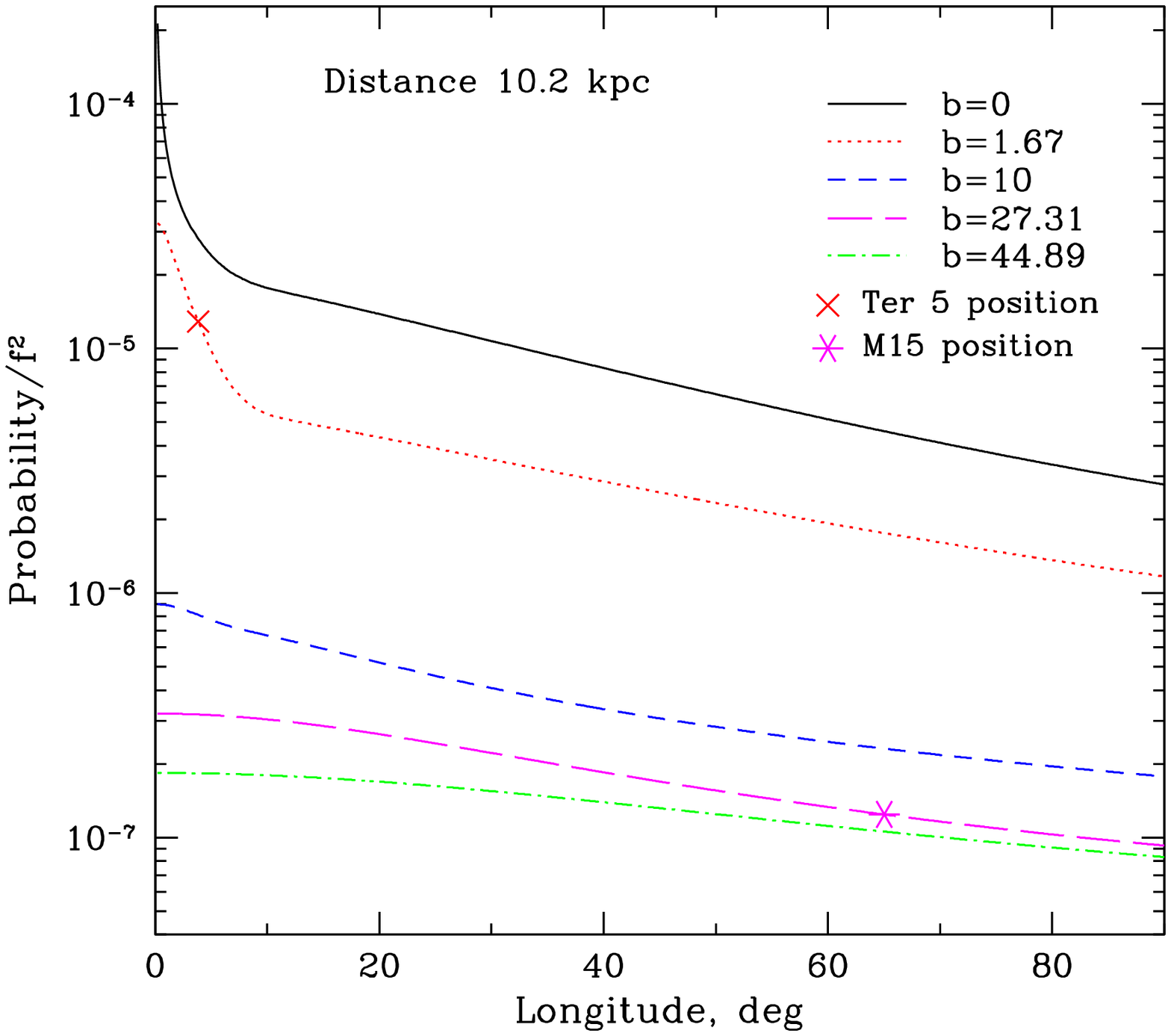}

\includegraphics[width=8.5cm,bb=35 265 545 680,clip]{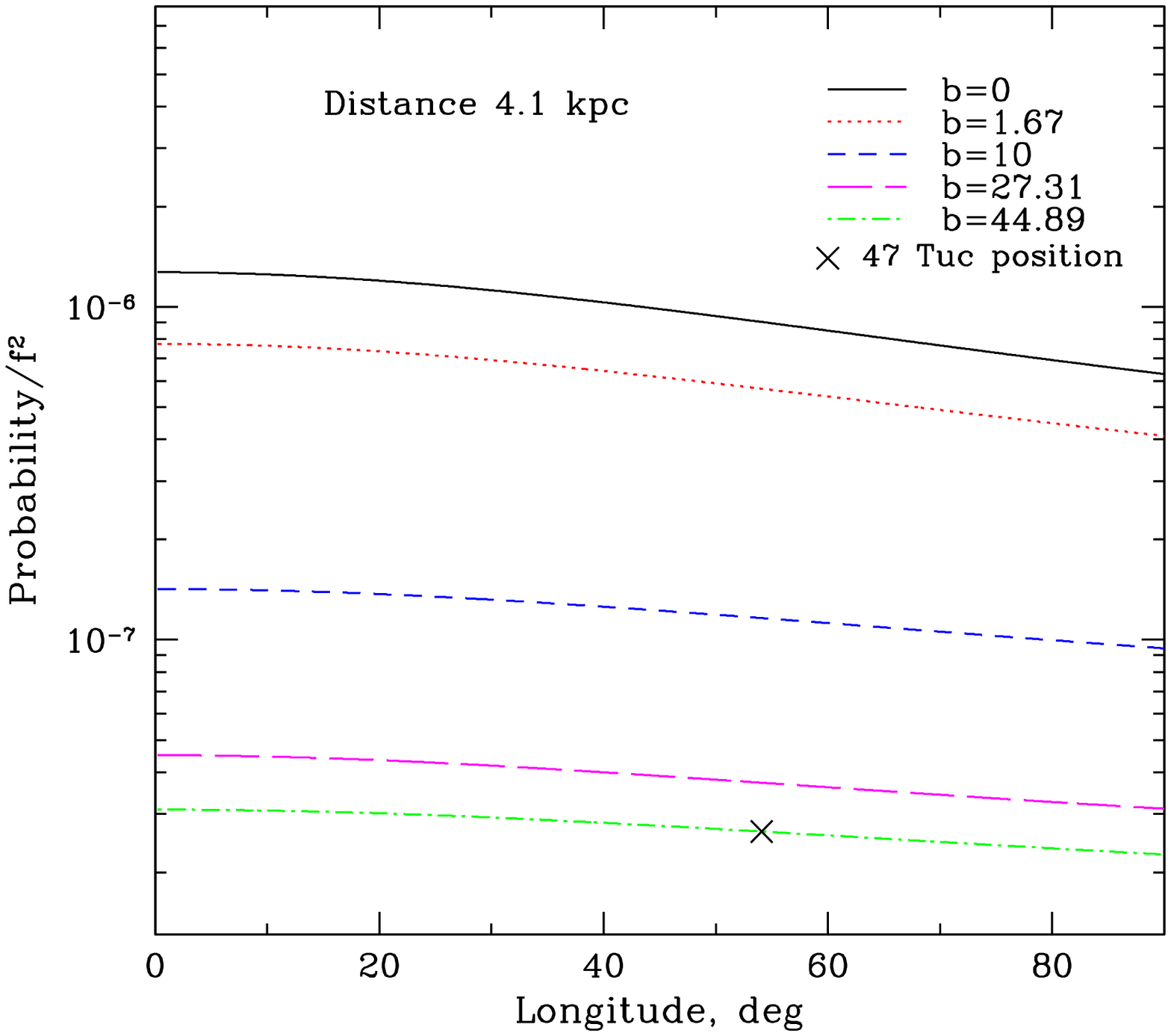}

\includegraphics[width=8.5cm,bb=35 215 545 680,clip]{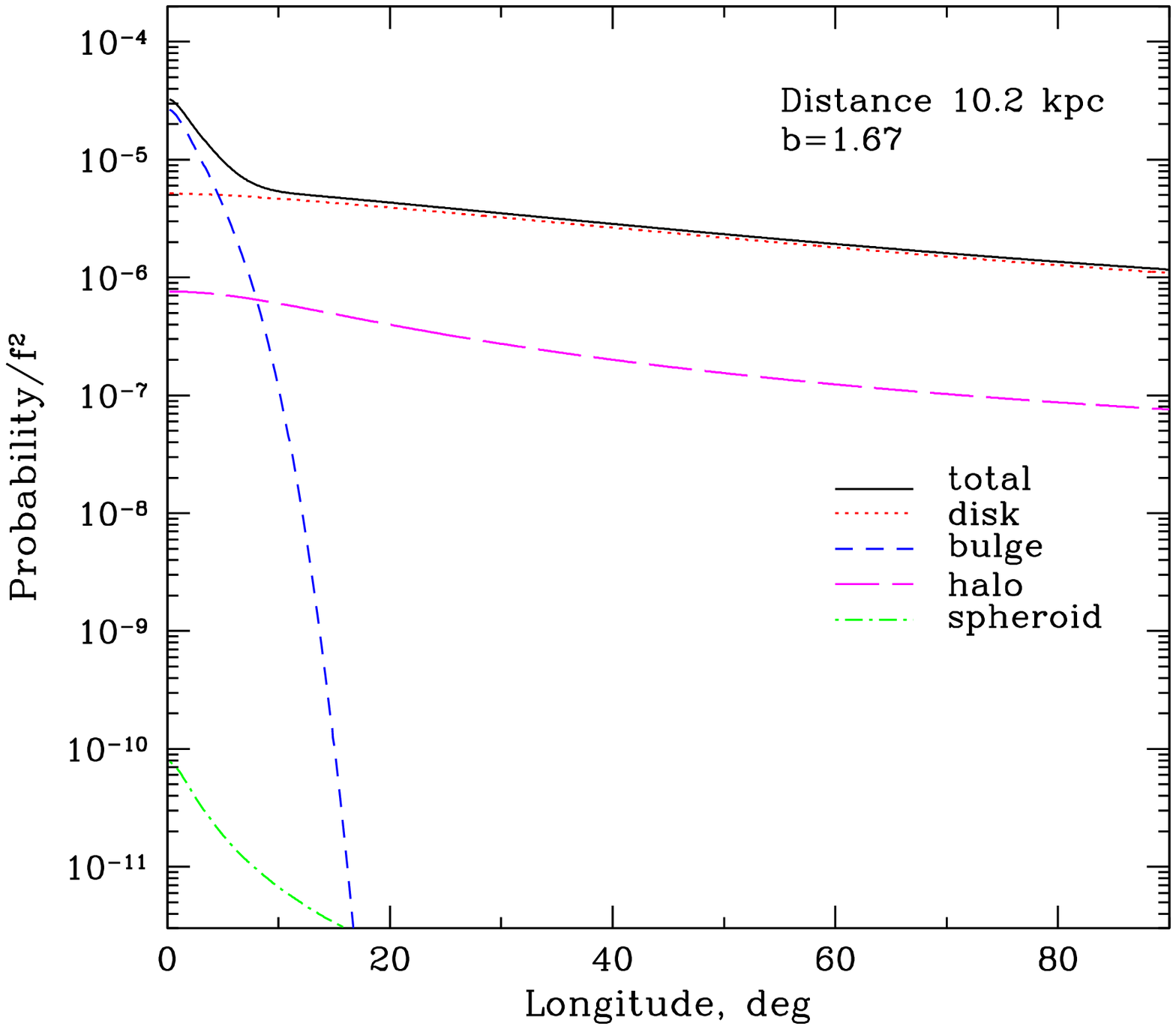}
\caption{\rm Same as Fig. 6 for the BS model of the Galaxy.}

\end{center}
\end{figure*}

\begin{figure}
\begin{center}
\includegraphics[width=12cm,bb=35 150 550 700]{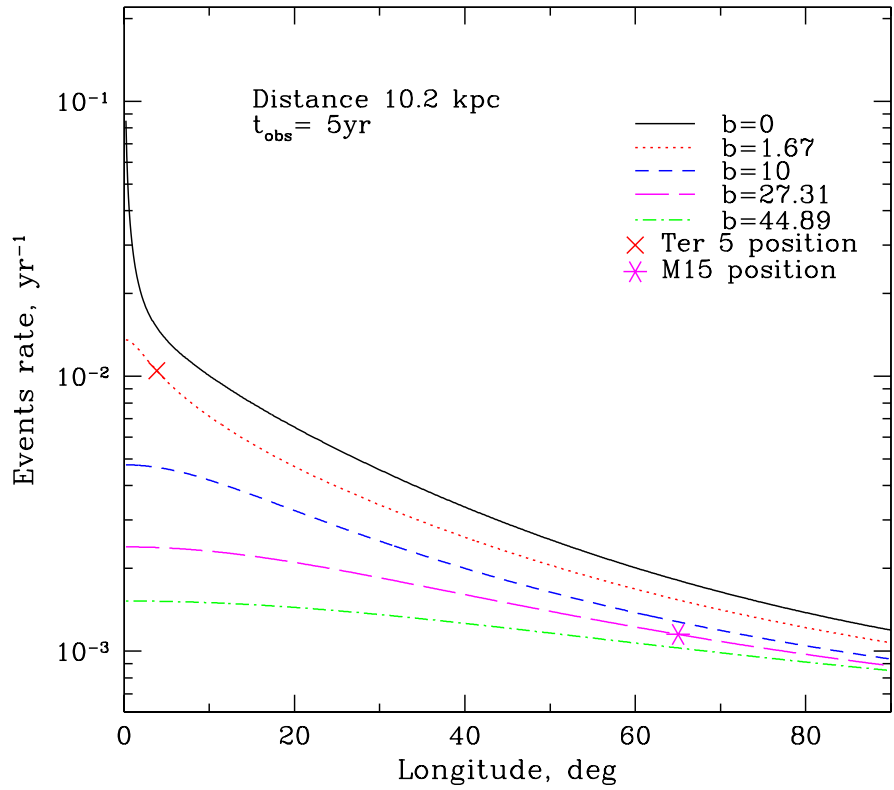}
\caption{\rm Event rate as a function of Galactic longitude $l$ and
latitude $b$ of a pulsar at a distance of 10.2 kpc. The positions of
the globular clusters Terzan 5 and M15 are indicated by the cross
and the asterisk, respectively. The DB model of the Galaxy is used.
}

\end{center}
\end{figure}

\begin{figure}
\begin{center}

\includegraphics[width=12cm,bb=35 150 550 700]{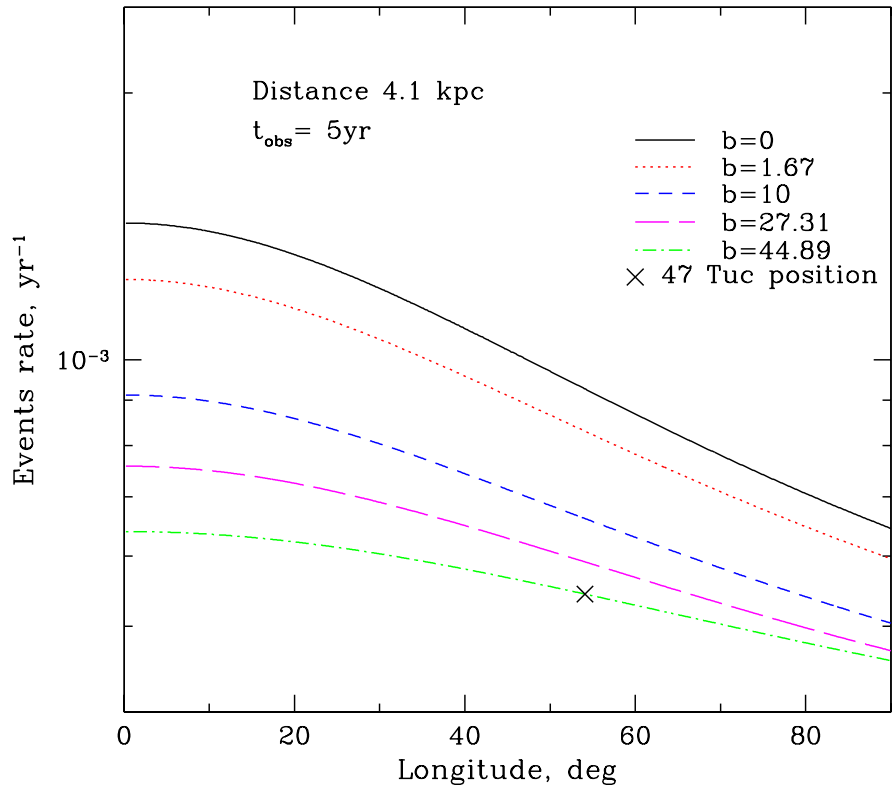}
\caption{\rm Event rate as a function of Galactic longitude $l$ and
latitude $b$ of a pulsar at a distance of 4.1 kpc. The position of
the globular cluster 47 Tuc is indicated by the cross. The DB model
of the Galaxy is used.  }

\end{center}
\end{figure}

\end{document}